# *Compact Variable-Gap Undulator with Hydraulic-Assist Driver.*

Alexander Temnykh[1*] and Ivan Temnykh [2]

[1]CHESS, Cornell University, Ithaca, NY 14850, USA;

[2] Pine Hollow Auto Diagnostics, Pennsylvania Furnace, PA 16865, USA

* Corresponding author, E-mail: abt6@cornell.edu

## Abstract

Following successful proof-of-concept experiments, we constructed compact varying-gap undulator with Hydraulic-Assist Driver. The $1.5\ m$ long undulator features a pure permanent magnet structure, period of $28.4\ mm$ and a minimum gap of $6.75\ mm$. The bench test of the device indicated excellent mechanical and magnetic performances. It demonstrated $\leq 3°\ RMS$ optical phase errors in a wide range of gap variation; $\sim 1.0 \times 10^{-4}$ (or better) repeatability in normalized $K_{und}$ ($\Delta K/K$) from one gap variation to another; $\sim 1.0 \times 10^{-4}\ RMS$ of $\Delta K/K$ long term (~24hrs) stability. In special mode, when the gap was varied in correlation with undulator temperature, undulator demonstrated $\Delta K/K$ stability at the level $5.3 \times 10^{-5}\ RMS$ while the undulator temperature change was $\sim 0.4\ °C$.

In this paper, we discuss the critical design components and criteria required for undulator mechanical stability and magnetic performance. We also describe our Lab-View-based control system governing the Hydraulic-Assist Driver. Finally, we present the test setup and results.

## 1. Introduction.

Presently, Cornell High Energy Synchrotron Source (CHESS) operates eight compact adjustable-phase undulators named Cornell Compact Undulators (CCUs). They have a $6.75\ mm$ constant gap, and the magnetic field is controlled by varying the phase between two magnet arrays [1].

The CCUs are superior to conventional varying-gap devices in several aspects: they are much more compact, lighter, more cost-efficient, and easy to build and operate. But in two characteristics they fall behind. First, at low field (or low $K_{und}$) operation, adjustable phase undulators have a trend to increase the phase errors, resulting in degradation of high order harmonics radiation. This trend does not affect regular operation, when experiments require maximum photon flux and therefore high $K_{und}$. However, for some experiments that require low SR power and therefore a low $K_{und}$, this trend can be undesirable. The second unwanted characteristic is the inability to open the gap during storage ring start-up or tuning. A small gap raises the risk of demagnetization caused by radiation (induced by particles lost from the electron/positron beam). Overcoming these two undesirable features of CCUs was the main motivation for the development of a compact undulator with a varying gap.

The compact varying-gap undulator described in this paper was constructed by adding a gap-opening mechanism based on the Hydraulic-Assist Driver [2] to one of the existing CCUs [1], with some structural modifications. To distinguish the varying-gap design from a regular CCU, it was named the Adjustable Gap Cornell Compact Undulator (AG CCU).

In sections 2, 3 and 4 we discuss the AG CCU design providing mechanical stability and therefore magnetic performance.   In Section 5 we describe the control system for the Hydraulic-Assist Driver. Finally, in Sections 6-8 we present results of bench testing the AG CCU. The test was focused on validation of the design as well as on short-term and long-term $K_{und}$ stability and reproducibility. There we also describe a mode of operation which greatly (by a factor 20) reduces sensitivity of the undulator field (i.e. of $K_{und}$ ) to ambient temperature.  This mode of operation is, probably, specific for AG CCU and cannot be applied to conventional gap adjustable undulators.

## 2. General design overview

The AG CCU is ~1.5 $m$ long, ~250 $mm$ wide, ~360 $mm$ high and weighs ~200 $kg$. It has a 6.75 $mm$ minimum gap, and a PPM structure with a 28.4 $mm$ period.  The AG CCU on the test bench and engineering model views are presented in Fig. 1. More detailed views are given in the Appendix.

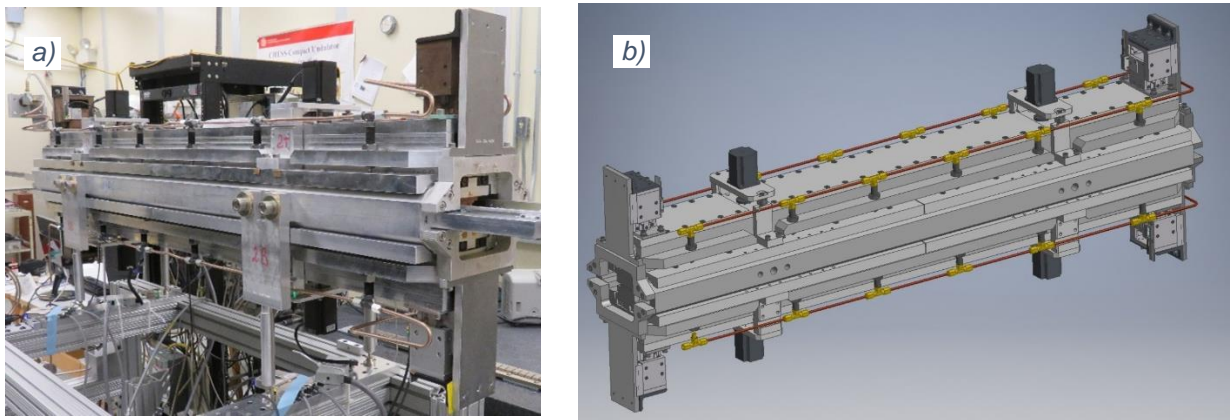

Figure 1.  a) AG CCU on a testing bench. b) Engineering model

The device consists of a stationary frame and two movable girders. Frame (Fig. 1A in Appendix) is made from two aluminum C-shape side plates connected to each other with bridges at the ends. Four mechanical drivers are attached to side plates at "minimum sag" points [5]. Drivers have "π" shape frames. They employ simple linear actuators and load cells providing real-time monitoring of a driver's load.  Movable girders (Fig. 2A in Appendix) consist of PM arrays, structural stiffness elements, and 12 (per girder) miniature hydraulic cylinders connected in series by a single hydraulic line. Cylinder bodies are fixed to the girders; their rods are pushing against frame with force controlled by pressure in hydraulic line. In operation, the hydraulic force compensates for most (~90%) of the magnetic forces and mechanical drivers are used only for accurate positioning of the girders. To restrict the girder's motion to only vertical direction, slider assemblies (Fig. 3A in Appendix) are placed at the girder's ends.

## 3. Key components

During design and construction of the AG CCU, special attention was paid to the following critical components:

## 3.1. Hydraulic System

AG CCU hydraulic system is similar in design to that which was developed in ref [2].

A key requirement was that the hydraulic system should be able to compensate magnetic forces at minimum gap. Knowing peak field at minimum gap, $B_0 = 0.95\ T$, undulator length $L = 1.5\ m$, pole width $W = 27.8\ mm$ and using expression for magnetic force:

$$F_{mag}[N] = \frac{B_0^2[T]}{4\mu_0} \times L[m] \times W[m]$$

we estimated the maximum magnetic forces as $7,881\ N$ or $1,770\ lbf$. That defined required hydraulic system load capacity.

Miniature hydraulic cylinders that we used in the project are model 20-01-07 from Vektek. They have 5/8'' DIA threaded body, pistons with 0.11 sq. inch (~71 sq. mm) area, 0.75 inch (19mm) stroke, 550 lbf (2,450 N) load capacity and rated for 5000 Psi ($3.4 \times 10^7\ Pa$) pressure. Based on these specs, one can suggest that 4 cylinders with ~4000 Psi pressure in the line would be enough to compensate for the magnetic force. For two reasons we employed 12 cylinders per girder. First, with larger number of cylinders the distance between them becomes smaller. That drastically reduces deformation of the girder, see Section 4. Second, for 12 cylinders per girder, required pressure is just 1341 Psi. For that range, a large variety of hardware and fittings is available in local automotive or hardware stores.

All 24 cylinders, 12 on upper girder and 12 on lower, were connected to a single hydraulic line equipped with a pressure transducer (WIKA, Type A-10/0 - 3000 psi).

As a hydraulic pressure source, we used a hydraulic ram pump (Pittsburgh Hydraulics 10,000-PSI model) driven by a linear actuator (Kollmorgen, model EC2).

## 3.2. Mechanical Drivers

AG CCU has four mechanical drivers, two for each girder, located at "minimum sag" position. Drivers consist of $\pi$-shape frames made of three 0.5in thick aluminum plates, linear actuator from LIN ENGINEERING rated up to 330 lb.- force (1430 N) and miniature in-line load cell (model LCM300 from Futek) ranged for +-250 lb.-force full scale. The load cells were used for constant monitoring of the force applied by actuators to girders. It should be noted that due to driver geometry, see Fig. 2, the cell reading is 50% of the actuator load.

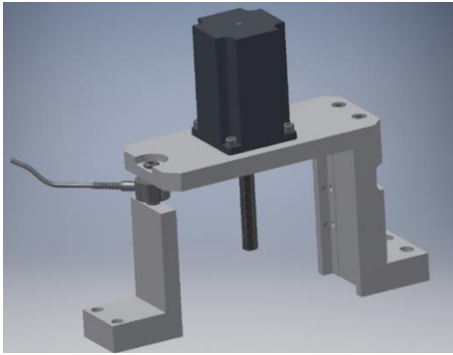

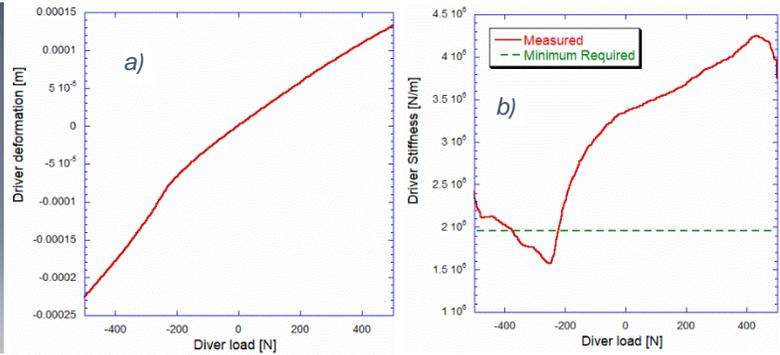

Figure 2. Mechanical driver consisting of "π" shape frame linear actuator and load cell.

Figure 3. Mechanical Driver Test Results. a) - Driver deformation measured as a function of load, b) Driver stiffness obtained from the deformation data. Dashed line indicates the required minimum.

According to criteria developed in Section 4, to prevent mechanical instability related to specific of operation of hydraulic system, stiffness of the mechanical drivers should be greater than $1.96 \times 10^6 \ N/m$.

In the course of prototyping, we have tested the mechanical drivers for stiffness. First, we measured deformation under varying load, then extracted the stiffness. Test results are depicted in Fig. 3. Results indicated that at positive load when linear actuator is pulling the nut, the driver stiffness satisfies criteria for the undulator mechanical stability (see Section 4.2). The strong dependence of stiffness on load is probably related to actuator design and preload scheme.

### 3.3. Guiding Assemblies and Encoders

To avoid horizontal displacement and rolling of the girders, we attached guiding assemblies (Fig. 3A) to both ends of the girders. The assemblies employed preloaded crossed roller rail sets. This preload provided stiffness needed to prevent girder's roll instability.

Four Renishaw TONiC encoders with 0.5 micron resolution are used to monitor the upper and lower array positions with respect to the frame. Encoders are installed on guiding assemblies (Fig 3A) at both ends of the girders.

## 4. Mechanical design consideration

To ensure AG CCU performance, RMS of phase errors originated from the girder's deformation causing magnetic field distortion should be smaller than 3°. That implies limit on acceptable deformation which depends on mechanical design. In this section, we present results of the AG CCU engineering model deformation analysis under various conditions. This analysis included calculation of phase errors. In addition, we evaluate criteria on required stiffness of mechanical drivers and guiding assemblies. This criteria should be satisfied to provide mechanical stability of the system with hydraulic components.

## 4.1. Girder deformation consideration.

To understand girder deformation, we performed both analytical and FEM analysis. For analytical, we used simplified girder model and well-developed Euler–Bernoulli beam theory [3]. For FEM, we analyzed realistic engineering girder model using ANSYS software [6].

### 4.1.1. Analytical model

Consider simplified girder model under magnetic force load $q(z)$ compensated by hydraulic cylinders uniformly distributed along girder with period $d$. The girder deformation $w(z)$ can be described by the Euler-Bernoulli static beam equation [3]:

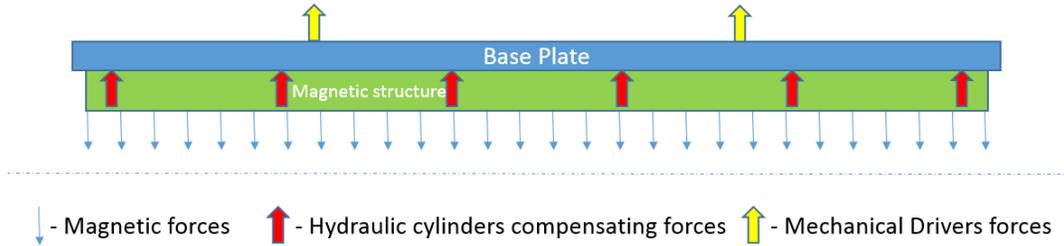

*Figure 4. Simplified girder model*

$$EI \frac{d^4 w(z)}{dz^4} = q(z)$$

Where "$E$" is the elastic modulus and "$I$" is the second moment of girder cross-section area.

If we exclude ends, i.e., assume infinite girder length, and uniform magnetic force load, $q(z) = q_m$, $q_m$ is a load per unit of length, equation can be easily resolved. Solution $w(z)$ will be periodical function and can be expressed as:

$$w(z) = \frac{q_m}{24EI}(z-d)^2 z^2 \; ; 0 < z < d$$

The maximum and average deformations through the period will be:

$$w_{max} = \frac{q_m}{384EI} d^4 \; ; w_{average} = \frac{q_m}{720EI} d^4$$

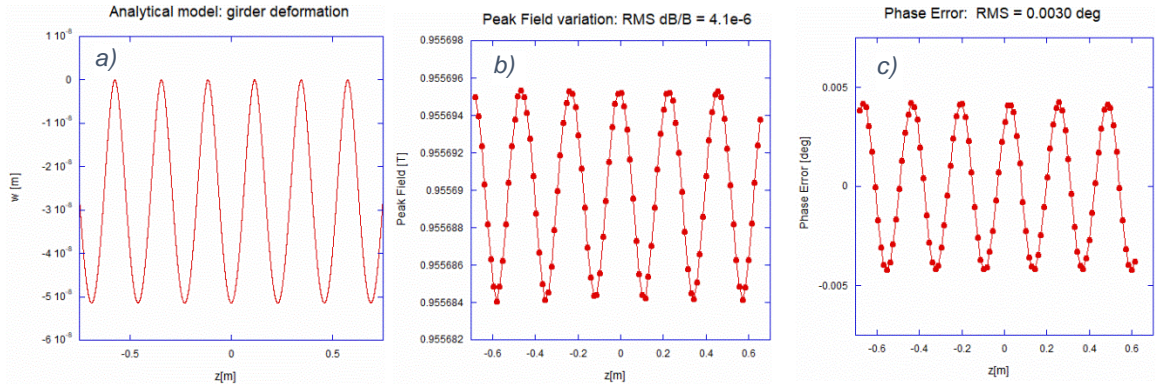

*Figure 5. Analytical model prediction (no ends included): a) girder deformation; b) peak field variation; c) phase errors*

Note, that the amplitude of the girder deformation has strong dependence on distance between compensation cylinders. It is proportional to the distance in fourth power.

For the model parameters: $d = 0.230\ m$; $q_m = 5212\ N/m$; second girder cross-section (Fig. 7) area integral $I = 1.04 \times 10^{-5}\ m^4$ and aluminum elastic modulus $E = 7.1 \times 10^{10}\ Pa$ it gives:

$$w_{max} = -0.5 \times 10^{-7}\ m;\ \ w_{average} = -0.27 \times 10^{-7}\ m$$

Fig. 5 shows plotted $w(z)$, peak field variation and phase errors. Normalized peak field variation ($dB/B\ RMS = 4.1 \times 10^{-6}$) and phase errors ($PhaseErr\ RMS = 3 \times 10^{-3}$) are negligible.

According to the concept, the hydraulic cylinders should compensate for 90% (or more) of the magnetic forces load. The rest should be taken by a two mechanical drivers located at "*minimum sag*" points [5]. These drivers are critical for mechanical stability of the device as well.

Let's evaluate the girder deformation resulted from combination of the remaining magnetic forces (those not compensated for by hydraulic system) and forces applied by drivers. In addition, we will take into account dependence of magnetic forces on gap, i.e. dependence on the girder deformation. The latter introduces non-linearity in the model making pure analytical solution non-trivial.

Equation describing girder deformation under above assumption will have a form:

$$EI \frac{d^4 w(z)}{dz^4} = \alpha q_m + q'_m \times w(z) + f_{driver}(z)$$

Here: $\alpha$ is the fraction of a magnetic forces not compensated by hydraulic cylinders (we assume $\alpha = 0.1$); term $q'_m \times w(z)$ describes magnetic force variation due to gap change resulted from girder deformation; $f_{driver}(z)$ is a force applied by mechanical drives. From experiment and from 3D magnetic field simulation we found $q'_m = -2.38 \times 10^6\ \frac{N}{m^2}$. The driver forces, $f_{driver}(z)$, should satisfy balance condition:

$$\int_0^{L_{und}} \left( \alpha q_m + q'_m \times w(z) + f_{driver}(z) \right) dz = 0$$

Here $L_{und} = 1.512\ m$ is undulator length. Assuming the position of two drivers satisfies condition for minimum sag, see [5], i.e. the distance between them $D = 0.5536 \times L_{und} = 0.8370\ m$ and they locate symmetrically in respect to undulator center we can find solution.

Because, as mentioned above, equation for girder deformation became nonlinear and not easy to resolve analytically, we found solutions numerically using *Ordinary Differential Equations* (ODE) solver in "MATLAB". Plots in Fig.6 show results.

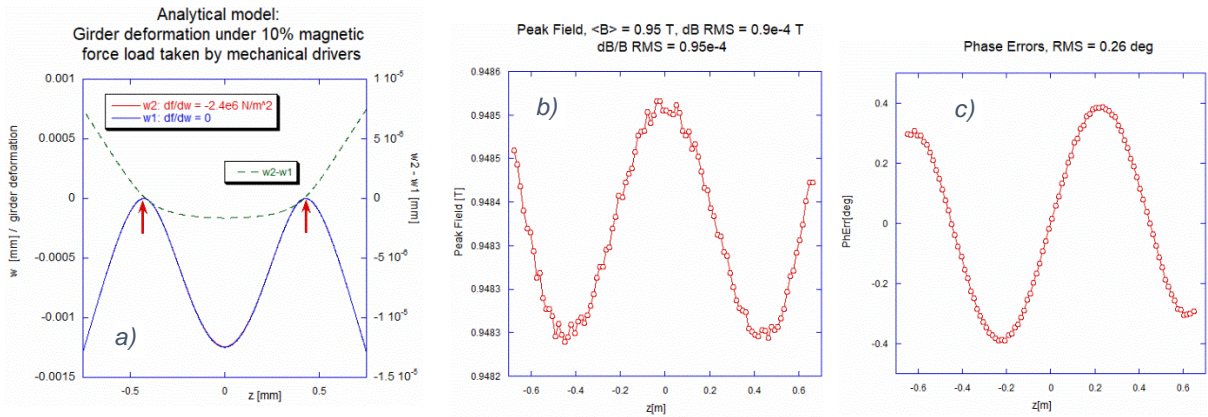

Figure 6. ODE analysis results. a) girder deformation under 10% of magnetic force load balanced by mechanical drivers. Red arrows show drivers location, w(z) RMS = 0.0043mm; b) peak field variation resulted from the girder deformation; c) phase errors resulted from the deformation.

Girder deformation data are on the left. Two plots showing solutions for the "flat" magnetic forces, $q'_m = 0$, and for more realistic case $q'_m \neq 0$, are practically overlapping. Both indicate approximately $\sim 10^{-3}$ $mm$ deformation and not more than 1% difference, plotted by dashed line with scale on the right. The small effect of the field variation with gap can by understood from comparison between terms $\alpha q_m = 521$ $N/m$ and $q'_m \times w(z) \sim 2.38$ $N/m$ for $w(z) \sim 10^{-3}$ $mm$. Note the small effect of the field variation with gap on girder deformation makes eligible the FEM analysis with loads independent of deformation. Magnetic peak field variation resulted from this girder deformation (plot in the middle) and phase errors (plot on the right) are quite acceptable.

### 4.1.2. Girder engineering model FEM study

Because of limitation of analytical approach we carried out FEM analysis of more realistic girder engineering model using software ANSYS. The model had real girder geometry with all actual details (Fig 7).

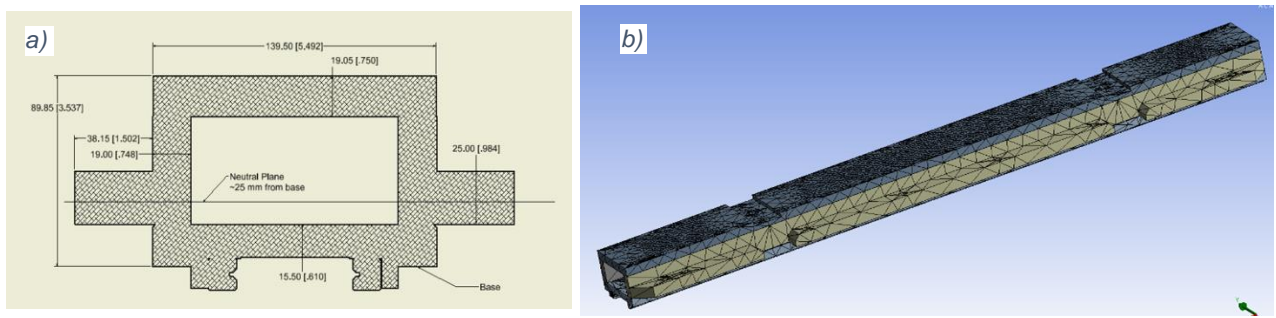

Figure7. a) girder cross-section with dimensions in mm and inches as alternative units. B) girder engineering model used for FEM analysis.

Result of ANSYS analysis of the girder deformation under magnetic force load fully compensated by hydraulic cylinders is shown in Fig.8. Here, plot "a" presents 3D model directional deformation. The deformation profile along the girder is shown by solid line on the plot "b". It indicates $\frac{-0.8\times 10^{-6}}{+4.0\times 10^{-6}}\ m$ displacement from the straight line with biggest at the ends. This is more than 10 times larger than predicted by simplified analytical model which did not take into account end effects. Peak field data plotted by dashed line on plot "b" revealed $dB/B\ RMS = 2.0 \times 10^{-4}$. Phase errors $RMS = 0.21°$, see plot "c". Although, girder deformation seems to be rather large, peak field variation and phase errors are acceptable.

Table 1 summarizes other simulation results. Case #1 is reproducing analytical model evaluated in the second part of section 4.1.1. Here we applied 10% (788 $N$) of magnetic force to the girder and balanced this force by a two mechanical drivers located at "minimum sag" points. In comparison with analytical model results, FEM analysis gave ~50% larger deformation, variation of peak field and phase errors. The difference between analytical and FEM study can be attributed to the used models. While in analytical study we assumed solid model with uniform cross-section, see Fig. 7a, in FEM analysis the model had many intrusions. That reduced girder rigidity and lead to larger deformation.

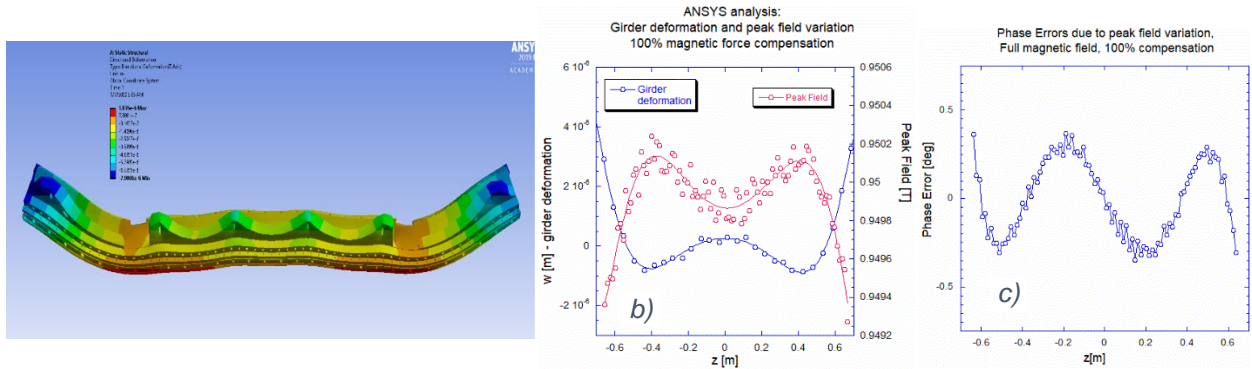

Figure 8. ANSYS simulation for the girder engineering model under magnetic forces 100% compensated by hydraulic cylinders. a) girder model directional (vertical) deformation. b) vertical deformation along the girder, RMS 1.7e-6m, and peak field variation caused by the deformation, RMS dB/B =2.0x10$^{-4}$. c) phase errors along the girder, RMS = 0.21º.

| Case # | Magnetic Force | Hydraulic compensating force / % | Mechanical drivers load | Girder deformation RMS [m] | Peak field variation dB/B RMS | Phase Error RMS |
|---|---|---|---|---|---|---|
| 1 | $-788.1\ N$ | 0 | 788 $N$ | $0.63 \times 10^{-6}$ | $1.47 \times 10^{-4}$ | 0.36 ° |
| 2 | $-7881\ N$ | $\frac{7093\ N}{90\ \%}$ | 788 $N$ | $1.2 \times 10^{-6}$ | $1.4 \times 10^{-4}$ | 0.26 ° |
| 3 | $-7881\ N$ | $\frac{7881\ N}{100\ \%}$ | 0 $N$ | $1.7 \times 10^{-6}$ | $2.0 \times 10^{-4}$ | 0.21 ° |
| 4 | $-7881\ N$ | $\frac{8669\ N}{110\ \%}$ | $-788\ N$ | $2.5 \times 10^{-6}$ | $3.8 \times 10^{-4}$ | 0.56 ° |

Table 1. Girder Deformation, peak field variation and phase errors for various level of magnetic force compensation.

Cases #2, 3 and 4 reflect designed range of operation. #2 and #3 are for extreme conditions when hydraulic system is under-compensating magnetic forces by 10 % and when it is over-compensating by 10%. #3 is for nominal situation when hydraulic system fully compensates for magnetic forces with no forces acting on mechanical driver. For the latter, the girder deformation, as well as peak field variation and phase error are plotted in Fig.8. Comparing results we can state that while under-compensation or full compensation would be preferable, the girder deformation, peak field variation and phase errors are all acceptable in the entire range.

## 4.2. Mechanical stability consideration

The operation of the AG CCU hydraulic system required detailed consideration. In our design, to move girder we add or remove oil from the system. In static state, when girder is not moving, the oil volume should be constant. It turns out that the constant oil volume is a necessary but not sufficient condition for girder stable position, since hydraulic cylinders are interconnected and oil can flow between them. The constant volume constraints average girder position, but not a pitch and roll. Because magnetic forces increase as the gap decreases, without additional constraints these forces will cause instability of pitching and rolling motion. We employed mechanical drivers and guiding assemblies with proper stiffness to add constraints and stabilize system.

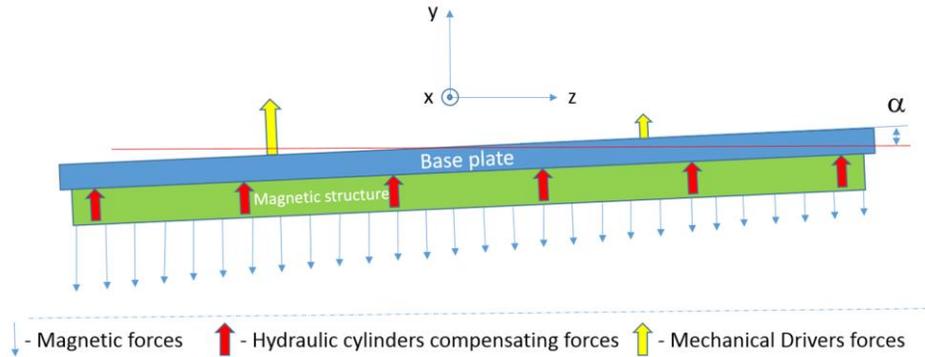

Figure 9. Schematic view of girder with pitch

Let's develop criteria for the required driver stiffness. Assume the girder was pitched by a small angle $\alpha$ as shown on Fig 9. The pitch will generate torque (moment of force) consisting of two terms. First term, caused by magnetic forces, can be estimated in the following way. The pitch generates gap ($\Delta y$) and magnetic forces ($\Delta f_m$) variation along undulator:

$$\Delta y = \alpha \times z;$$

$$\Delta f_m(z) = q'_m \times \Delta y = \alpha \times q'_m \times z$$

The force variation results in the torque:

$$\tau_m = \int_{-L_{und}/2}^{L_{und}/2} z \times \Delta f_m(z)\, dz = \alpha q'_m \int_{-L_{und}/2}^{L_{und}/2} z^2 dz = \alpha \frac{q'_m L_{und}^3}{12}$$

The second term is due to mechanical driver's reaction forces. It will be:

$$\tau_{driver} = D \times F_{driver} = D \times k_{driver} \times \frac{\alpha D}{2} = -\alpha \frac{k_{driver} D^2}{2}$$

Here $k_{driver}$ is a mechanical driver stiffness. To have mechanical stability the total torque should be negative for positive $\alpha$, i.e.

$$\frac{k_{driver} D^2}{2} > \frac{q'_m L_{und}^3}{12}; \quad k_{driver} > \frac{q'_m L_{und}^3}{6D^2}$$

For our design ($q'_m = -2.38 \times 10^6 \frac{N}{m^2}, L_{und} = 1.512 \, m, D = 0.553 \times L_{und} = 0.836m$), it gives:

$$k_{drive} > 1.96 \times 10^6 \, N/m$$

In Section 3.2, we compared results of the driver stiffness test with criteria developed above. Comparison indicated satisfactory for the chosen driver design.

The girder rolling was constrained by the guiding assemblies coupling the girders with the undulator frame. Special attention was paid to the stiffness of the assemblies.

The bench test of the AG CCU described in the following sections demonstrated good mechanical stability of the device and validated results of the design analysis.

## 5. Instrumentation and control system

Monitoring part of the AG CCU control system consisted of 10 channels: one hydraulic pressure transducer (WIKA, Type A-10 / 0 - 3000 psi), four load cells (FUTEK, model LCM300), four high accuracy linear encoders (Renishaw, TONiC™ T103x) and one thermistor (YSI 44005). Signals from pressure transducer, load cells and thermistor were digitized by high accuracy digital multimeters. Galil Motion Control DMC_2182 controller was used to read linear encoders.

Controlling part had 5 output channels governing stepping motors. These are a linear actuator (Kollmorgen, model EC2) driving hydraulic ram pump and four linear actuators (Lin Engineering, Inc.) used in mechanical drivers.

To operate the whole system, we developed a program using LabView software. Basically, the program executed two independent feedback loops at a 1Hz rate.

In the first loop, the program was monitoring the upper and bottom girders position by reading the encoder signals while moving mechanical drivers in small steps toward the target. Before each step, the program was checking the load cells signals, i.e. mechanical drivers load. The motion was activated only if the load was "low". The "low load" condition was satisfied only when the force created by the hydraulic cylinders compensated 90% of magnetic forces between girders.

The second loop controlled the hydraulic pressure. In this loop, the program read encoders, calculated the ID gap and calculated the hydraulic pressure needed for ~95% compensation of the magnetic forces. The desired pressure was compared with the actual (measured by the pressure transducer) and, if the desired pressure was higher, the program activated the hydraulic pump to add oil into the system.

If the desired pressure was lower than the actual, the program just cycled idly until the actual pressure was reduced due to leaks.

The program Interface (Fig. 10) displays in real time all monitoring parameters, such as position of the girders, hydraulic line pressure, undulator temperature etc. Interface also allows control of the undulator gap, hydraulic pressure (hydraulic pump), the mechanical drivers and also provides an interface with other programs. Other programs were used to cycle or ramp up/down the gap, activate Hall probe scans to measure magnetic field profile, record various parameters and so on.

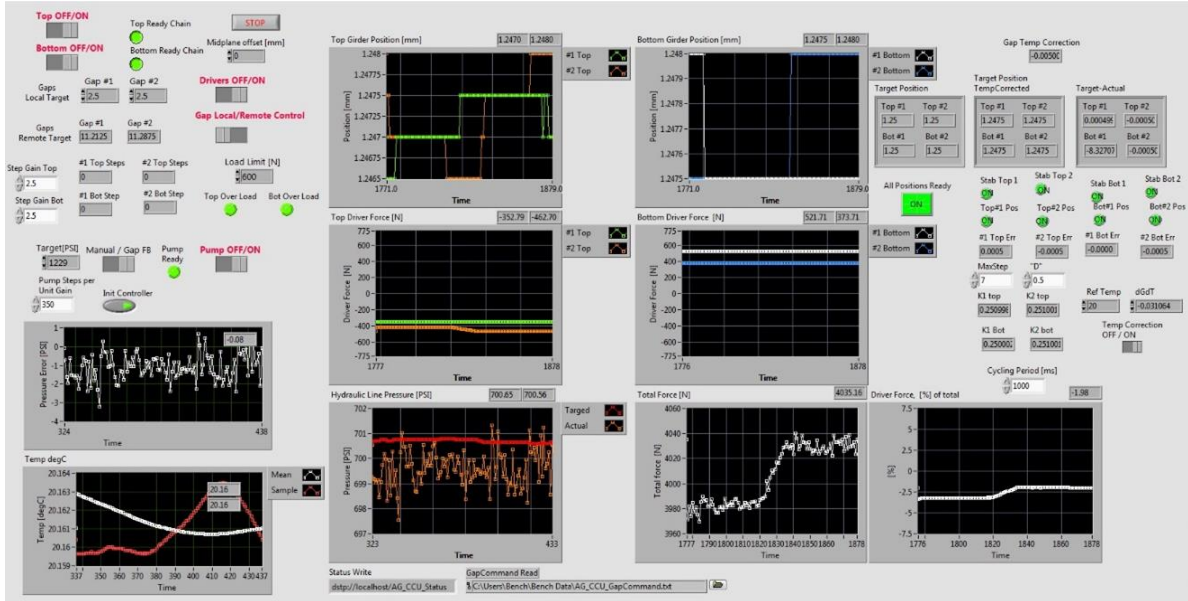

*Figure 10. Screen shot of the interface of the program controlling undulator.*

## 6. Test setup

Magnetic field measurement setup was similar to the setup developed in Ref. [6] for closed undulator

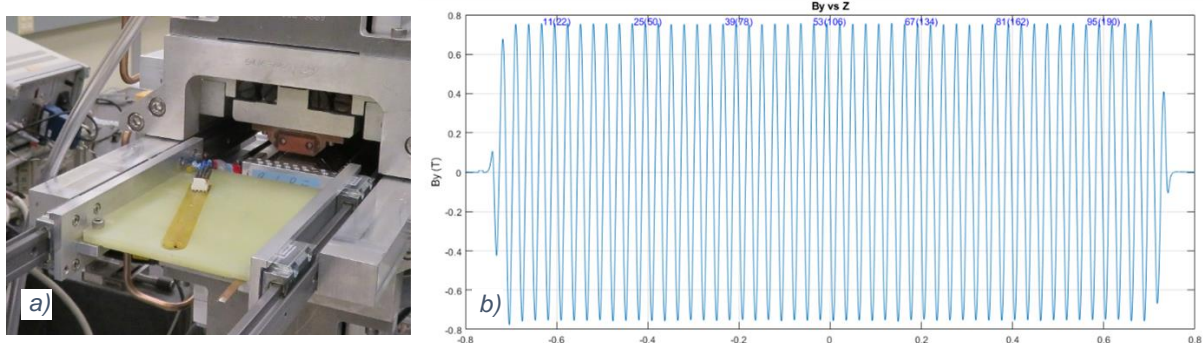

*Figure 11. a) Magnetic field measurement setup. Shown are guiding rails, sliders from IGUS and G10 plate with attached Hall sensor. b) – measured magnetic field profile at 7mm undulator gap.*

structure.

The setup consisted of two non-magnetic guiding rails from IGUS attached to the frame in middle plane inside of undulator, rectangular G10 plate with four sliders moved along the rails by two long (~2.3m) carbon rods and Hall sensor (Lakeshore HGT-2101) attached to the plate. Carbon rods were coupled to magnetic field measurement bench carriage. Hall sensor was connected to "LakeShore 455 DSP Gaussmeter " by four 34 AWG wires. Magnetic field measurements were taken "on the fly". One scan comprised of approximately 9,000 measurements and required approximately 2 minutes of time.  For analysis we used modified package of MatLab scripts originally developed by Zachary R. Wolf in SLAC in 2012.

## 7. Test results

To evaluate mechanical and magnetic performance of AG CCU, we carried out a number of tests addressing various topics. First, we tested mechanical stability focusing on the repeatability of girder positioning and possible girder deformation. Test results are presented in the following "Mechanical performance" subsection.  Then, when we studied long term magnetic field stability, we noticed a strong correlation between undulator temperature variation and magnetic field change.  After introducing of gap correction for the temperature change, undulator demonstrated normalized field long term stability at $\sim 1 \times 10^{-4}$ in $\sim \pm 0.5°C$  temperature range. These results are presented in "Magnetic performance" subsection.

### 7.1. Mechanical performance

Mechanical performance of the AG CCU was evaluated through analysis of magnetic field measurement data.

To evaluate accuracy of the girder positioning we varied undulator gap and measured dependence of $K_{und}$ on the gap. This dependence was used for accuracy estimation.

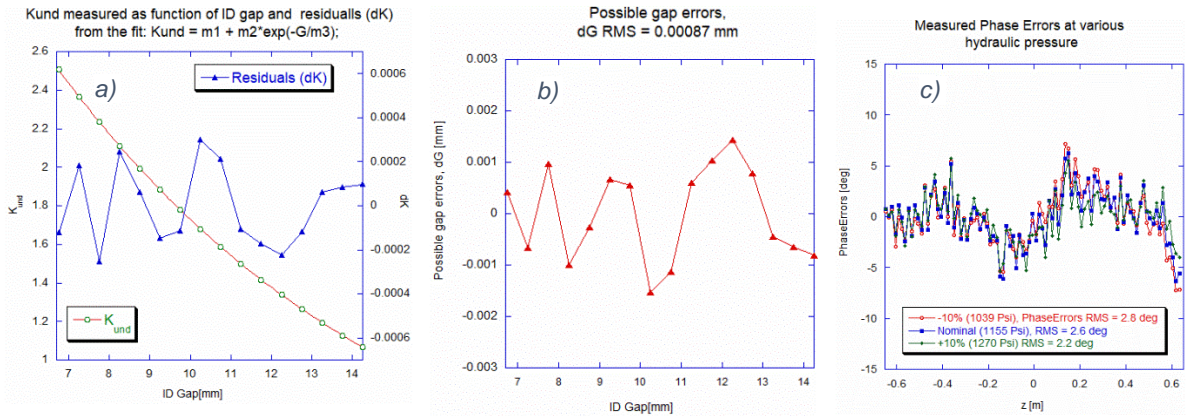

*Figure 11. a) $K_{und}$ as function of undulator gap (left scale) and residuals from the fit (right scale); b) ID gap errors required for explanation of residuals seen on plot "a)";  c) Phase Errors at minimum gap measured at various hydraulic pressure corresponding to 90%, 100% and 110% of magnetic force compensation.*

Two plots on Fig. 11a represent $K_{und}$ versus undulator gap (left scale) and fitting residuals (right scale). For the fit we used formula: $K_{und} = m_1 + m_2 \times exp(-G[mm]/m_3)$.  The last-squares fit yields $m_1 = 0.0141; m_2 = 5.4207; m_3 = 8.6882\ mm$ with $1.7 \times 10^{-4}$ residuals $(dK)$ RMS.  These residuals can have two origins. There can be contribution from Hall probe signal drift and/or contribution from the girder positioning errors causing gap errors. Let's assume stable Hall probe and calculate gap errors required for explanation of observed $dK$. In this way, we will find the upper limit. Taking derivation of the fitting formula we find: $\frac{dK}{dG} = -\frac{m_2}{m_3}exp(-G/m_3)$ . It translates to: $dG = -\frac{m_3}{m_2}exp(G/m_3) \times dK$.

Plot in Fig. 11b shows $dG$ data. It is in $\pm 1.5 \times 10^{-3}$ $mm$ range with $0.87 \times 10^{-3}$ $mm$ RMS. Knowing that the gap errors resulted from both upper and bottom girders position errors, we can estimate an upper limit on RMS of possible errors in girder position as $\frac{0.87 \times 10^{-3}}{\sqrt{2}} = 0.61 \times 10^{-3}$ $mm$. Taking into account $0.5 \times 10^{-3}$ $mm$ resolution of the encoders we used to control positions, we can state that the estimated limit is very reasonable, meaning the accuracy of the girder positioning is satisfactory.

In the next experiments we studied undulator performance under the load-simulating conditions considered in the second half of section 4.1.2. Fig. 11c represents phase errors data obtained at minimal ID gap and with various hydraulic pressures corresponding to 90%, 100% and 110% of magnetic force compensation. These conditions correspond to the cases #2, 3 and 4 presented in the Section 4.1.2 Table 1. The data indicates variation of RMS phase errors by ~0.6° while magnetic force compensation was changed from 90 to 110%. It is consistent with expectations (see numbers in Table 1) and is acceptable.

Small variation of phase errors, meaning small girder deformation, confirmed adequate rigidity of the girders and correct location of the mechanical drivers.

## 7.2. Magnetic performance

The operation of undulators as a synchrotron radiation source requires, in addition to small phase errors, the long-term stability of, $K_{und}$, as well as its high repeatability after gap variation. Because $K_{und}$ depends on the gap and in the gap controlling mechanism we used many innovations, we tested $K_{und}$ long term stability and repeatability with great attention.

The phase errors, $K_{und}$ long term stability and repeatability test results are presented below.

### 7.2.1. Phase Errors

The phase errors were measured as a function of gap. Results are presented in Fig. 12.

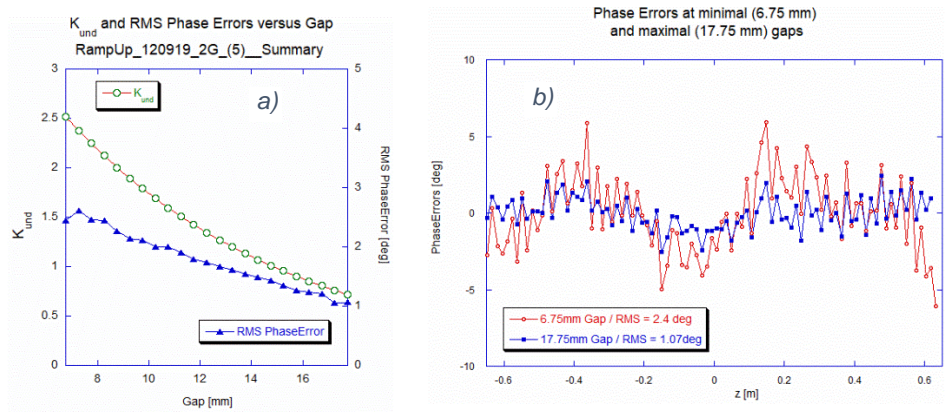

*Figure 12. Undulator Phase Errors. a) $K_{und}$ and phase errors RMS versus gap. b) Phase errors along undulator measured at minimal and maximal gaps.*

The data plotted on Fig. 12 a) indicate a 2.4° phase error RMS at minimal 6.75 $mm$ gap. As the gap is increased, the phase error decreases. At a 17.75 $mm$ gap, the phase error RMS is 1.1°. Both the phase error RMS and the dependence on gap, are typical for adjustable gap undulators. Fig. 12 b) shows phase errors along undulator at minimal and maximal gaps.

## 7.2.2. Undulator parameter long-term stability and its stabilizing with "TGap" correction.

Time record data are shown on Fig 12a and Fig. 12c. Data points were taken with 6.5 minute interval and included undulator temperature, gap and "K" parameter. To determine "K", undulator was scanned with

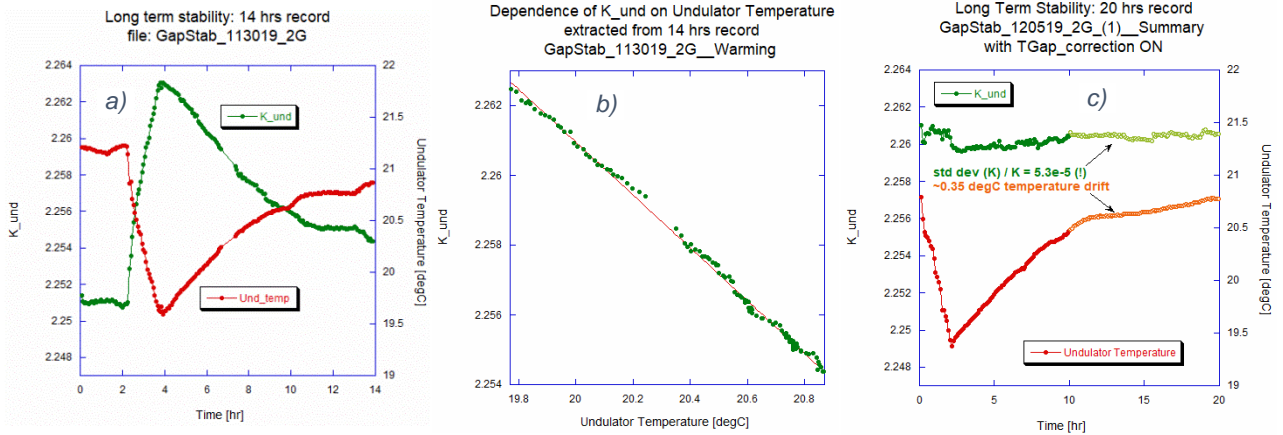

*Figure 13. Long term stability test. a) $K_{und}$ and undulator temperature 14 hour record for 6.75 mm constant undulator gap; b) observed dependence of $K_{und}$ on temperature, linear fit gives $(dK/K)/dT = -0.00335 \pm 1.4 \times 10^{-5}$ °C$^{-1}$; c) $K_{und}$ and undulator temperature record with "TGap" correction applied.*

Hall Probe, then magnetic field was analyzed and $K_{und}$ was extracted. Undulator temperature was measured with a thermistor attached to the frame.

The first 14 hour $K_{und}$ record (Fig. 12a) was taken at a constant 6.75 $mm$ gap. It shows $K_{und}$ variation from 2.2510 to 2.2629 ($\Delta K/K \sim 5.2 \times 10^{-3}$) correlated with undulator temperature change from 21.1 °C to 19.6 °C ($\Delta T \cong -1.6$°C). It should be noted the room where we tested undulator had no climate control. The ambient temperature was usually quite unstable. It depended on time the day, outside weather and so on and can be unpredictably changed in range $\pm 0.5$°C. After we noticed correlation between $K_{und}$ and undulator temperature in the previous experiments, to amplify the effect of $K_{und}$ change we used air conditioning to intentionally lower the room temperature during recording period.

On Fig.12b we plotted recorded data in form of $K_{und}$ dependence of undulator temperature. The plot reveals well-defined linear dependence and the linear fit:

$$K_{und} = K_0 \times \left(1 + \widetilde{m}_1 (T - T_{ref})\right)$$

yields $\widetilde{m}_1 = \frac{\left(\frac{dK}{K}\right)}{dT} = -0.003352 \pm 0.000014$ °C$^{-1}$, $K_0 = 2.2610 \pm 0.00002$ for $T_{ref} = 20$°C.

In the next experiment we varied gap to compensate $K_{und}$ change caused by temperature variation.

Let's find the gap change ($\Delta G$) needed for this compensation. Combining formulas used for fitting dependences $K_{und}$ versus "gap" and $K_{und}$ versus temperature, we can write an expression for $K_{und}$ variation when we have both, small temperature and gap changes:

$$\Delta K_{und} = K_0 \widetilde{m}_1 \Delta T - \frac{K_0}{m_3} \Delta G$$

To simplify expression we neglected the term $m_1$ in comparison with $m_2 \times exp(-G[mm]/m_3)$ in the dependence $K_{und}(G)$. To have $\Delta K_{und} = 0$, we should introduce temperature-to-gap ("TGap") correction:

$$\Delta G[mm] = \tilde{m}_1 m_3 \Delta T = -0.02912 \times \Delta T[°C]$$

The "19 hours" record data presented on Fig. 12c was taken with continuous "TGap" correction. In the loop controlling girder position, we added correction according to the above expression. As for $\Delta T$ we used the difference between 20°C reference temperature and measured. Similar to the previous, in the beginning of the record period we intentionally lowered undulator temperature by 1.6°C by switching on and off air-conditioning. The data indicates that "TGap" correction drastically improved $K_{und}$ stability. With "TGap" correction ON and 1.6°C temperature change, the observed Pk-Pk amplitude of normalized $K_{und}$ variation was $6.2 \times 10^{-4}$ with $RMS = 1.3 \times 10^{-4}$ over the record period. Without correction for similar temperature change it was 10 times larger. Peak-to-peak $(dK/K)$ variation was $4.9 \times 10^{-3}$ with $RMS = 1.6 \times 10^{-3}$.

"19 hours" record with "TGap" correction ON indicated the largest $K_{und}$ intrusion at moment 2 hour from the start when we cooling was stopped and undulator temperatures started to rise. The intrusion resulted from the small difference between the temperatures at PM blocks and thermistor locations. This difference depends on the temperature change rate, for slower change it is smaller, and for falling temperature the difference will be positive and it will be negative for rising temperature. That varying temperature difference affects precision of correction.

After "10 hour" point, undulator temperature slowly drifted up by ~0.35°C in the following 10 hours. Knowing $K_{und}$ dependence on temperature, Fig. 13b, one can expect normalized $K_{und}$ drift ~$0.003352°C^{-1} \times 0.35°C = 1.2 \times 10^{-3}$. In reality, due to the "TGap" correction, it was much smaller. Pt-Pt normalized $K_{und}$ variation was $2.6 \times 10^{-4}$ with RMS $0.53 \times 10^{-4}$. Long-term stability at this level will satisfy the most demanding applications.

### 7.2.3. Undulator parameter repeatability

Two evaluate $K_{und}$ repeatability we varied (ramped up) the gap from 6.75 to 17.75 mm with 0.25mm step and after each step measured $K_{und}$. The summary of five such ramps is presented in Fig. 14.

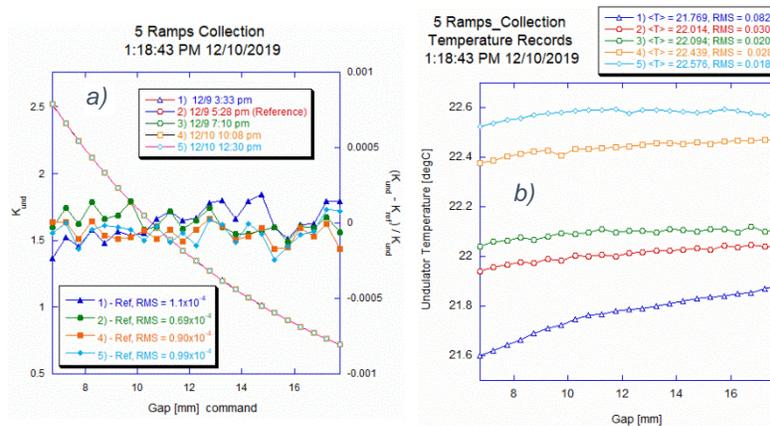

Figure 14. $K_{und}$ repeatability test. a) Five data sets. Open marks and left-hand scale indicate measured $K_{und}$ versus gap. Closed marks and right-hand scale show deviation from reference "#2)" measurement. b) Undulator temperature recorded in parallel with $K_{und}$ measurements.

Three ramp measurements #1, 2 and 3 were done in one day with 2-hour interval; measurements #4 and #5 were performed the next day. In repeatability evaluation we used one measurement (#2) as a reference and analyzed the difference between this measurement and others. The data plotted in Fig. 14a by a solid marks with left-hand scale is that normalized difference. For all measurements Pt-Pt variation of normalized $K_{und}$ is less than $4.0 \times 10^{-4}$ with $0.88 \times 10^{-4}$ RMS. These numbers indicate excellent demonstrated repeatability.

It should be noted that all measurements were done with "TGap" correction ON. The temperature record presented on Fig. 14b indicated that there was noticeable temperature drift during each ramp as well as between them. For example, the temperature difference between ramp #1 and #5 was ~1.0°C. Knowing how $K_{und}$ depends on temperature, see Fig. 13b, one would expect the normalized $K_{und}$ difference between ramp #1 and #5 at ~0.0033 level. In fact, due to "TGap" correction it was much lower at $1 \times 10^{-4}$ level.

## 8. Discussion and conclusion

Based on concept developed in ref [2], we built and tested the Adjustable Gap Cornell Compact Undulator (AG CCU), a $1.5\ m$ -long compact undulator, where the gap was adjusted using the developed Hydraulic-Assist Driver.

Great attention was paid to the design of undulator structural elements in order to minimize their deformation. The design was validated through both analytical models and FEM analysis. We also analyzed potential mechanical instability due to specifics of hydraulic system operation and defined criteria for resolving this issue. Mechanical components were constructed in accordance with these criteria to provide stable operation of the AG CCU.

The AG CCU demonstrated excellent *mechanical* performance: Positioning accuracy of the girders, i.e. gap accuracy, at $1 \times 10^{-3}\ mm\ RMS$ level; ~0.5° change of RMS phase errors under various conditions of hydraulic system operation. These results validated the mechanical design.

The AG CCU demonstrated excellent *magnetic* performance as well. Measurements showed: ~2.4° $RMS$ phase errors at minimum gap with $K_{und} = 2.521$ and smaller errors at larger gap; repeatability and long-term stability of normalized $K_{und}$ at $1 \times 10^{-4}$ level or better. This level of repeatability and long-term stability was achieved due to employing "TGap" correction: gap correction depending on undulator temperature.

Similar to the CCU described in [1], the new AG CCU is compact, lightweight, easy to fabricate and cost-efficient. The operation of the AG CCU is somewhat more complicated, requiring a number of transducers, encoders, and a control program to execute the precise gap adjustment. However, a wide range of programming tools and standard industrial instrumentation is readily available. A stand-alone controller, similar to an "automotive engine control module" and standardized set of hardware would bring down the cost and setup time even further.

Because the built AG CCU satisfies all operational requirements, there is a plan to use it in regular CHESS operation after a period of beam testing.

## Acknowledgment

The Work was supported by the NSF award DMR-1332208 and by the efforts from Pine Hollow Auto Diagnostics.

## References

[1] A. Temnykh *et al.*, Compact Undulator for Cornell High Energy Synchrotron Source, IEEE Transactions on Applied Superconductivity, Vol. 22, No. 3, June 2012.

[2] A. Temnykh, Ivan Temnykh and Eric Banta. Hydraulic-assist driver for compact insertion devices, Nuclear Inst. and Methods in Physics Research, A 917 (2019) 18–24.

[3] Euler–Bernoulli beam theory: https://en.wikipedia.org/wiki/Euler%E2%80%93Bernoulli_beam_theory

[4] ANSYS, Inc. Products 2019 R2

[5] Airy points:

https://en.wikipedia.org/wiki/Airy_points#Other_support_points_of_interest

[6] AIP Conference Proceedings 2054, 030037 (2019); https://doi.org/10.1063/1.5084600

# Appendix

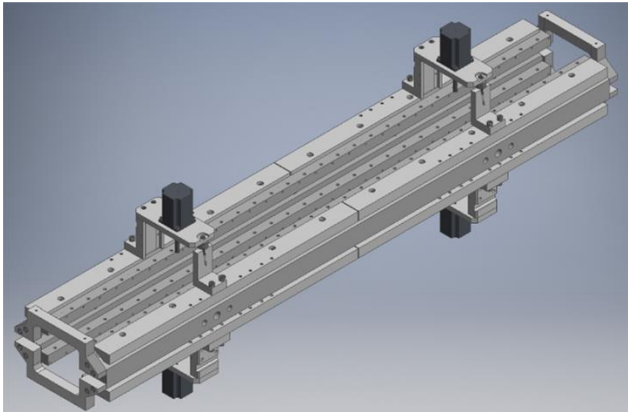

*Figure 1A. AG CCU stationary frame. Four mechanical drivers with linear actuators and load cells are attached to the frame at "minimum sag" points.*

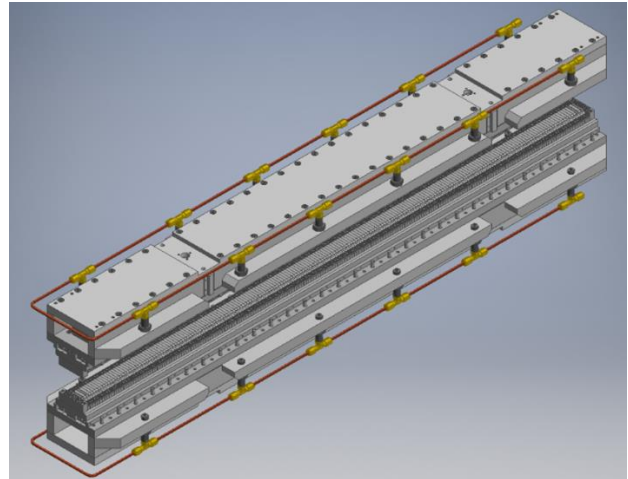

*Figure 2A. AG CCU Movable girders with attached miniature hydraulic cylinders (12 per girder rendered in black). Cylinders are connected in series by a hydraulic line.*

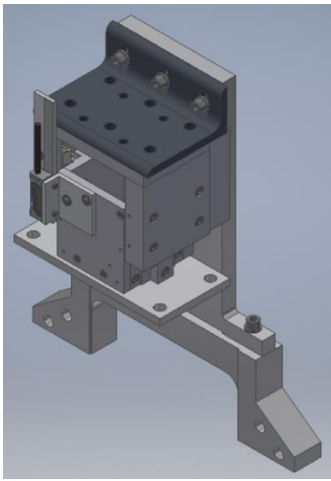

*Figure 3A. Guiding assembly providing link between frame and moving girders. They restricts girder motion to one degree of freedom.*